# Electron shelving induced lasing in cold bosonic atoms in optical lattice

Aranyabhuti Bhattacherjee[+]

Department of Physics, Atma Ram Sanatan Dharma College, University of Delhi (South Campus), Dhaula Kuan, New Delhi-110 021, India.[*]
Electronic address: bhattach@df.unipi.it

And

The Abdus Salam International Centre for Theoretical Physics, Trieste, Italy.

**Abstract**: We calculate the absorption spectrum for ultracold three-level helium atom in a $\Lambda$ configuration in an optical lattice. Our results show the possibility of lasing at certain points on the optical lattice which are capable of rendering one of the transitions metastable as compared to the other. A coherent control over the stimulated emission is possible using an axial magnetic field.

[*]: Address for Correspondence

# INTRODUCTION

Modern atomic physics has advanced primarily by using known physics to devise innovative techniques to better isolate and control the atomic system, and then exploiting this nearly ideal system to achieve higher precision and discover new physical phenomena. One of the most striking advances along these lines has been "Laser cooling and trapping of atoms and quantum manipulation". Recent spectacular advances in this field [1] provides an opportunity for novel investigation of the details of the interaction of light and matter and to probe concepts developed for condensed matter at totally different scales using optical lattice, where the defect free confining potential is created by light interference. These lattices can be made essentially free of defects with only moderate care in spatially filtering the laser beams to assure a single transverse mode structure. Furthermore, the shape of the potential is exactly known, and does not depend on atomic energy level scheme. Finally, the laser parameters can be varied, and the lattice vectors can be changed independently by redirecting the laser beams. These optical traps that have achieved kinetic temperatures as low as nanokelvins, have now created the ultimate physical systems thus far for precision spectroscopy, frequency standards, quantum computing and tests of fundamental physics. One such interesting concept of "electron shelving in optical lattice" was proposed by us recently[2] for a cold Helium atom in an optical lattice. We found that certain points on the optical lattice induce electron shelving in the helium system. These special points make one of the transition weak as compared to the other. Hence the weak transition becomes the metastable transition. We also found that the coherence between the two ground states of the helium atom responsible for population trapping also controls the photon statistics. A weak axial magnetic field was found to be a convenient tunable experimental parameter, which could efficiently control the ground state coherence and hence the photon statistics. This motivated us to look for the possibility of exploiting the long life time of the metastable state ( induced by the polarization gradient in the optical lattice) to create atomic inversion and at the same time have a coherent control over the stimulated transition probability to obtain gain using the weak axial magnetic field. Recently [3,4], it was proposed to utilize cold helium atoms in the Raman configuration as a UV-laser gain medium in the limit of photon recoil shift exceeding the natural atomic line width and the Doppler broadening.

# THE MODEL

Figure1 shows a Λ atomic system driven by two counterpropagating orthogonal plane-polarized laser fields with the same frequency $\omega_L$. For the experiments with $^4$He, $g_\pm$ are the two Zeeman sublevels m=$\pm$1 of the $2^3S_1$ state and $e_o$ is the m=0 Zeeman sublevel of $2^3P_1$. The light shifts of the Zeeman sublevels are position dependent, since the couplings (C.G. couplings) vary with polarization. The superposition of the two counter propagating orthogonally-polarized plane waves gives rise to a standing wave for which the polarization is space dependent. The polarization is alternatively circular $\sigma^-$ in Z=0, linear in Z=$\lambda$/8, circular $\sigma^+$ in Z=$\lambda$/4 etc., with a spatial periodicity $\lambda$/2. This polarization gradient leads to a periodic spatial modulation of the degenerate ground state Zeeman sublevels, which acts as a potential for the motion of the atomic centre of mass. An axial magnetic field provides the necessary coupling between the two degenerate ground states even for atoms with zero momentum. The Sisyphus cooling mechanism of fig.2 traps the atom in the laser induced optical potential well for blue detuning. The time evolution of the atomic density matrix ρ, in the Schrodinger picture, obeys

$$\frac{d\rho}{dt} = -\frac{i}{\hbar}[\hat{H}, \rho] + \left(\frac{d\rho}{dt}\right)_{sp} \qquad (1)$$

Here the total Hamiltonian H consists of an atomic Hamiltonian $H_A$, a laser – atom interaction Hamiltonian $H_{LA} = -\vec{D}.\vec{E}$, where $\vec{D}$ is the electric dipole moment operator, and the positive frequency part of the electric field $\vec{E}$ propagating along the z-axis is

$$\vec{E}^+(z,t) = \frac{E}{\sqrt{2}}[\vec{\varepsilon}_+ e^{i\pi/4}\cos(kz) + \vec{\varepsilon}_- e^{-i\pi/4}\sin(kz)] \qquad (2)$$

with $\vec{\varepsilon}_\mp = \mp(\vec{e}_x \pm i\vec{e}_y)/\sqrt{2}$, where $\vec{\varepsilon}_+$ and $\vec{\varepsilon}_-$ are the polarization unit vectors, a term $H_B$ corresponding to the shift in the energy due to an applied axial magnetic field B. Term $H_B$ corresponding to the shift in the energy due to an applied axial magnetic field $(\frac{d\rho}{dt})_{sp}$ is the damping term by spontaneous emission processes. We will work in a space spanned by $|i,p\rangle$, where $i$ is the index to the atomic energy level, and $p$ is the momentum eigenvalue of the atomic center-of-mass motion. In this space, the atomic Hamiltonian

$$H_A = \sum_i \int dp (\hbar\omega_i + \frac{p^2}{2m}) |i,p\rangle\langle p,i| \qquad (3)$$

is diagonalized. The interaction Hamiltonian, under the rotating-wave approximation, becomes

$$H_{LA} = -\int dp (\hbar\Omega_1 e^{-i\omega_L t} |e_o,p\rangle\langle p+\hbar k, g_+| + H.c) - \int dp (\hbar\Omega_2 e^{-i\omega_L t} |e_o,p\rangle\langle p-\hbar k, g_-| + H.c), \qquad (4)$$

where $\Omega_1 = \frac{\Omega}{2}\cos(kz)$ and $\Omega_2 = \frac{\Omega}{2}\sin(kz)$ are the usual Rabi equivalent space dependent field amplitudes. The Magnetic interaction term is

$$H_B = g\mu_B B \int dp (|p+\hbar k, g_+\rangle\langle g_+, p+\hbar k| - |p-\hbar k, g_-\rangle\langle g_-, p-\hbar k|), \qquad (5)$$

g is the gyromagnetic ratio for the $2^3S_1$ state, $\mu_B$ is the Bohr magneton = - 9.274 × 10$^{-24}$ J/T. (dρ/dt)$_{sp}$ is the damping due to spontaneous emission. Following the concept of momentum families [5-8], the fast oscillations with the optical frequency $\omega_L$ can be eliminated by taking

$$\tilde{\rho}_{e_o e_o}(t,p) = \rho_{e_o e_o}(t,p), \tilde{\rho}_{g_\pm g_\pm}(t,p) = \rho_{g_\pm g_\pm}(t, p\pm\hbar k, p\pm\hbar k), \tilde{\rho}_{g_+ g_-}(t,p) = \rho_{g_+ g_-}(t, p+\hbar k, p-\hbar k)$$

$$, \tilde{\rho}_{e_o g_\pm(t)}(t,p) = \rho_{e_o g_\pm}(t, p\pm\hbar k)\exp(i\omega_L t).$$

We obtain the generalized optical Bloch equations as:

$$\dot{\tilde{\rho}}_{e_o e_o}(t,p) = i\frac{\Omega}{2}[\cos(kz)\{\tilde{\rho}_{e_o g_+}(p) - \tilde{\rho}_{g_+ e_o}(p)\} - \sin(kz)\{\tilde{\rho}_{g_- e_o}(p) - \tilde{\rho}_{e_o g_-}(p)\}] - 2(\Gamma_{g_- g_-} + \Gamma_{g_+ g_+})\tilde{\rho}_{e_o e_o}(p)$$

$$\dot{\tilde{\rho}}_{g_+ g_+}(t,p) = i\frac{\Omega}{2}\cos(kz)[\tilde{\rho}_{g_+ e_o}(p) - \tilde{\rho}_{e_o g_+}(p)] + 2\Gamma_{g_+ g_+}\int_{-\hbar k}^{+\hbar k} N(q)\tilde{\rho}_{e_o e_o}(p + \hbar k + q)dq$$

$$\dot{\tilde{\rho}}_{g_- g_-}(t,p) = i\frac{\Omega}{2}\sin(kz)[\tilde{\rho}_{g_- e_o}(p) - \tilde{\rho}_{e_o g_-}(p)] + 2\Gamma_{g_- g_-}\int_{-\hbar k}^{+\hbar k} N(q)\tilde{\rho}_{e_o e_o}(p - \hbar k + q)dq$$

$$\dot{\tilde{\rho}}_{g_+ g_-}(t,p) = i\frac{\Omega}{2}[\sin(kz)\tilde{\rho}_{g_+ e_o}(p) - \cos(kz)\tilde{\rho}_{e_o g_-}(p)] - (2i\beta - \Gamma_{g_+ g_-})\tilde{\rho}_{g_+ g_-}(p)$$

$$\dot{\tilde{\rho}}_{e_o g_+}(t,p) = i(\delta + \omega_R + \beta)\tilde{\rho}_{e_o g_+}(p) + i\frac{\Omega}{2}[\cos(kz)\{\tilde{\rho}_{e_o e_o}(p) - \tilde{\rho}_{g_+ g_+}(p)\} - \sin(kz)\tilde{\rho}_{g_- g_+}(p)] - \Gamma_{g_+ g_+}\tilde{\rho}_{e_o g_+}(p)$$

$$\dot{\tilde{\rho}}_{g_- e_o}(t,p) = -i(\delta + \omega_R - \beta)\tilde{\rho}_{g_- e_o}(p) - i\frac{\Omega}{2}[\sin(kz)\{\tilde{\rho}_{e_o e_o}(p) - \tilde{\rho}_{g_- g_-}(p)\} - \cos(kz)\tilde{\rho}_{g_- g_+}(p)] - \Gamma_{g_- g_-}\tilde{\rho}_{g_- e_o}(p)$$
(6)

where $\Omega$ is the Rabi frequency, $\delta$ $(=\omega_L - \omega_A)$ is the detuning, $\omega_R = \hbar k^2/2m$ is the recoil frequency shift and $\beta = (kp/m + gB\mu_B/\hbar)$ is the Doppler shift associated with the velocity p/m in absence of the axial magnetic field. When $\beta \neq 0$ the energies of $|g_-\rangle$ and $|g_+\rangle$ differ by 2$\beta$. $N(q) = \frac{3}{4\hbar k}[1 - (\frac{q}{\hbar k})^2]$ is the distribution of spontaneous emission for a linearly polarized light. N(q) is normalized according to

$$\int_{-\hbar k}^{+\hbar k} N(q)dq = 1 \qquad (7)$$

$\Gamma_{g_+ g_-}$ can be viewed as a dephasing time between the ground levels due to collisions. The condition necessary for electron shelving is that one of the driving fields should be weak as compared to the other [9]. Consider for example points on the optical lattice where $\sin^2(kz) \gg \cos^2(kz)$, this necessarily puts $\Gamma_{g+g+} \ll \Gamma_{g-g-}$. This condition makes the $\sigma_-$ polarized field, which drives the $|e_o\rangle \leftrightarrow \langle g_+|$ transition weak as compared to the $\sigma_+$ polarized field, which drives the $|e_o\rangle \leftrightarrow \langle g_-|$ transition. Hence $|e_o\rangle \leftrightarrow \langle g_+|$ becomes the metastable transition and during this time the atom is predominantly in the metastable state (electron shelving). Dark periods (absence of resonance fluorescence) is a signature of electron shelving. This occurs at points near $z = n\lambda/4$, i.e. where the polarization is either $\sigma^+ or \sigma^-$ since here only one of the transition is more favored [2]. From experimental view $z = n\lambda/4$ are the points corresponding to bottom of the potential well where the atoms tend to accumulate once their kinetic energy is sufficiently small. As the atoms move along the optical lattice, it alternatively encounters points with $\sigma^+, linear, \sigma^-$ polarization. Points with linear polarization drive both the transitions equally. At points where $Sin^2(kz) = 0 (or Cos^2(kz) = 0)$ the atom behaves as a two level system since the light is entirely $\sigma_+ polarized (\sin^2(kz) = 0)$ or entirely $\sigma_- polarized (\cos^2(kz) = 0)$ and hence drives only one transition.

## NUMERICAL RESULTS AND DISCUSSION

We will take $\sin^2(kz) \gg \cos^2(kz)$ as an example to study our system. We monitor the rate of change of the number of laser photons in the $|e_o\rangle \leftrightarrow \langle g_+|$ transition, which is:

$$\frac{d\langle n_{g_+}\rangle}{dt} = -W_{abs}\tilde{\rho}_{g_+g_+}(p) + W_{em}[\tilde{\rho}_{g_-g_-}(p) + \tilde{\rho}_{e_oe_o}(p)] \qquad (8)$$

where $W_{abs}$ and $W_{em}$ are the stimulated absorption and emission rates which are calculated following reference [10]. The system exhibits gain for simple analytical expressions for $W_{abs}$ and $W_{em}$ for $\int_{-\hbar k}^{+\hbar k} N(q)\tilde{\rho}_{e_oe_o}(p + \hbar k + q)dq = \tilde{\rho}_{e_oe_o}(p)$ and for closed family of momentum which satisfies the steady state population conservation relation $\tilde{\rho}_{g_-g_-}(p) + \tilde{\rho}_{g_+g_+}(p) + \tilde{\rho}_{e_oe_o}(p) = 1$.

$$W_{abs} = \frac{\Omega^2 \cos^2(kz)}{4\Delta^2}[16\beta^2(\Gamma_{g_+g_+} + \Gamma_{g_-g_-}) + \Gamma_{g_+g_-}[\frac{\Omega^2 \sin^2(kz)}{4} + \Gamma_{g_+g_-}(\Gamma_{g_+g_+} + \Gamma_{g_-g_-})]] \quad (9)$$

$$W_{em} = \{\frac{\frac{\Omega}{4\Delta^2}\cos^2(kz)}{(\Gamma_{g_+g_+} + \Gamma_{g_-g_-})[(\Gamma_{g_+g_+} + \Gamma_{g_-g_-}) + 4(\delta' - \beta)^2] + 2(\Gamma_{g_+g_+} + \Gamma_{g_-g_-})\frac{\Omega^2 \sin^2(kz)}{4}}\}$$

$$\times \{\frac{\Omega^2 \sin^2(kz)}{4}[(\Gamma_{g_-g_-} + \Gamma_{g_+g_+})(\Gamma_{g_-g_-} + \Gamma_{g_+g_-})(\frac{\Omega^2 \sin^2(kz)}{4} + \Gamma_{g_+g_-}(\Gamma_{g_+g_+} + \Gamma_{g_-g_-})) +$$
$$16\beta^2 \Gamma_{g_+g_+}(\Gamma_{g_+g_+} + \Gamma_{g_-g_-}) + 4(\delta' + \beta)^2 \Gamma_{g_+g_-}\Gamma_{g_-g_-} - 8\beta\Gamma_{g_+g_-}(\delta' + \beta)\}$$

(10)

Where

$$|\Delta|^2 = [\frac{\Omega^2 \sin^2(kz)}{4} + \Gamma_{g_+g_-}(\Gamma_{g_+g_+} + \Gamma_{g_-g_-}) - 8\beta(\delta' + \beta)]^2 + [2\Gamma_{g_+g_-}(\delta' + \beta) + 4\beta(\Gamma_{g_+g_+} + \Gamma_{g_-g_-})]^2$$

(11)

The steady state solutions of the Bloch equations under the conditions $\sin^2(kz) \gg \cos^2(kz)$ and $\Gamma_{g+g+} \ll \Gamma_{g-g-}$ is

$$\tilde{\rho}_{g_-g_-} = \frac{16\beta^2[\frac{\Omega^2}{4}\sin^2(kz) + 4\Gamma_{g_-g_-} + 4(\delta' - \beta)^2] + \frac{\Omega^2}{16}\cos^2(kz) + 16\frac{\Omega^2}{4}\cos^2(kz)\beta(\delta' - \beta)}{N}$$

(12)

$$\tilde{\rho}_{e_o e_0} = \frac{16\beta^2 \frac{\Omega^2}{4}\sin^2(kz)}{N} \quad (13)$$

$$\tilde{\rho}_{g_+g_+} = \frac{\frac{\Omega^4}{16}\sin^2(kz)}{N} \quad (14)$$

with  $N = 32\beta^2(\frac{\Omega^2}{4}\sin^2(kz) + 2\Gamma_{g_-g_-} + 4(\delta' - \beta)^2 + \frac{\Omega^2}{16} + 16\frac{\Omega^2}{4}\cos^2(kz)\beta(\delta' - \beta)$

(15)

The general shape of the absorption spectrum as a function of β and the detuning is shown in the 3-dimensional surface plot ( fig.3.)The characteristic feature that one can immediately observe is a pronounced asymmetry in the absorption profile with respect to the detuning parameter $\delta'$. Absorption is seen to occur at blueshifted frequencies while stimulated emission can be observed for red detuning. Another remarkable feature one notes is that the spectrum changes from absorption to gain as β increases and vanishes for higher values of β. For β = 0, the atom shows absorption of the probe photons. As β increases (for a fixed value of the detuning parameter) the atom begins to emit. A maxima in the emission spectrum is obtained for a certain value of the parameter β. Further increase in β causes the spectrum to vanish. A transparent description of this problem can be reached in the coupled $|\Psi_c\rangle$ and noncoupled $|\Psi_{NC}\rangle$ state basis [11]. This basis is composed of the following two orthogonal linear combinations of $|g_-\rangle$ and $|g_+\rangle$.

$$|\Psi_C\rangle = \frac{1}{\sqrt{2}}\{\cos(kz)|g_+\rangle + \sin(kz)|g_-\rangle\} \quad (16)$$

$$|\Psi_{NC}\rangle = \frac{1}{\sqrt{2}}\{\sin(kz)|g_+\rangle - \cos(kz)|g_-\rangle\} \quad (17)$$

The state $|\Psi_{NC}\rangle$ is always decoupled from the laser field for any value of β:

$$\langle e_o|V|\Psi_{NC}\rangle = 0. \quad (18)$$

On the other hand, the atomic motion and/or the applied weak axial magnetic field induces a coupling between $|\Psi_c\rangle$ and $|\Psi_{NC}\rangle$,

$$\langle \Psi_C|H_B|\Psi_{NC}\rangle = -g\mu_B\beta \quad (19)$$

Finally, the laser field couples $|\Psi_c\rangle$ to the excited state,

$$\langle e_o|V|\Psi_C\rangle = \frac{\hbar\Omega}{\sqrt{2}}\exp(-i\omega_L t). \qquad (20)$$

From equations (16) and (17), we find that for points where $\sin^2(kz) \gg \cos^2(kz)$, the weight of $|g_-\rangle$ in $|\Psi_c\rangle$ is much larger than the weight of $|g_+\rangle$. The conclusions are reversed for $|\Psi_{NC}\rangle$. Consequently $|\Psi_{NC}\rangle \approx \frac{|g_+\rangle}{\sqrt{2}}$ and $|\Psi_c\rangle \approx \frac{|g_-\rangle}{\sqrt{2}}$. In the absence of coupling (i.e $\beta=0$) between the $|\Psi_c\rangle$ and the $|\Psi_{NC}\rangle$ states, the upper level occupation probability tends to zero so that absorptive processes dominate in this limit, and finds its explanation in the fact that the uncoupled state is stable against laser excitation. In general the spontaneous decay rate from the non-coupled state $\Gamma_{NC}$ ( $=\Gamma_{g_+g_+}$ for $\sin^2(kz) \gg \cos^2(kz)$) is a function of $\beta$ and is very small when $\beta$ is small, and vanishes for $\beta=0$. This means that an atom, which is put in $|\Psi_{NC}\rangle \approx \frac{|g_+\rangle}{\sqrt{2}}$ at time t = 0, can remain there for a very long time (if $\beta$ is small enough) on the order of $\Gamma_{NC}^{-1}$. Once the atom is trapped in the uncoupled state, it is unavailable for excitation to the excited state but can only be populated via spontaneous or stimulated emission from the upper level. This can happen only for a finite value of the coupling parameter. Let us consider the atom to be initially in the $|g_-\rangle$ state at a point on the optical lattice where $\sin^2(kz) \gg \cos^2(kz)$. At this point the atom will be optically pumped to the excited state by the strong $\sigma_+$ polarized field. The transition $|g_-\rangle \rightarrow |e_o\rangle \rightarrow |g_+\rangle$ is maximum where $|g_-\rangle$ energy is highest. Once the atom is in the excited state, it has a finite rate to jump either to the weak ground state $|g_+\rangle$ or the strong ground state $|g_-\rangle$. Interaction with the vacuum fluctuations will induce the atom to drop to the $|g_+\rangle$ state (which forms a potential well at that point) via spontaneous emission. This process results in a loss of photons in the $|e_o\rangle \rightarrow |g_+\rangle$ transition. We identify two mechanisms by which the atom will exhibit gain. Now (from the $|g_+\rangle$ potential well) when the atom climbs uphill, its kinetic energy is transferred into potential energy by stimulated emission process. Atomic momentum is therefore transferred (as gain ) to the laser photons on the $|e_o\rangle \leftrightarrow \langle g_+|$ transition. This process of stimulated emission is operative as long as the atomic momentum is sufficiently high to convert kinetic energy into potential energy. In the absence the external magnetic field, this process of stimulated emission ceases as soon as the atom is trapped in the $|g_+\rangle$ potential as a result of insufficient momentum. The second contribution to the gain comes from the axial magnetic field which helps to restore the ground state coherence even in the absence of atomic momentum. Consider, for instance after a spontaneous decay the atom is trapped in $|\Psi_{NC}\rangle \approx \frac{|g_+\rangle}{\sqrt{2}}$. The coupling generated by the magnetic field transfers the atom to the $|\Psi_c\rangle \approx \frac{|g_-\rangle}{\sqrt{2}}$ ( i.e. to the potential hill of $|g_-\rangle$ ). Starting from $|g_-\rangle$, the atom can absorb a $\sigma^+$ photon in the k$\varepsilon_z$ wave, jump into the excited state, then make a stimulated emission of a $\sigma^-$ photon in the -k$\varepsilon_z$ wave, which brings it back into the $|g_+\rangle$ state. The rate of this transition is maximum only for a certain optimum value of the magnetic field. Population in the excited state also assumes a maximum value for this same value of the coupling parameter(figure not shown). For still larger values of β, (larger than the Rabi frequencies and the spontaneous emission rate of the excited state), the atom switching between the lower levels take place at a very small but finite occupation probability of the upper level. This explains the practically zero emission for high values of β.

## CONCLUSIONS

We have demonstrated that using a three level Λ scheme, metastable cold helium atoms can be used as a gain medium in an optical lattice. We took advantage of the metastable state induced by the polarization gradient and the asymmetry in the atomic absorption and emission spectrum. Gain is obtained at points on the optical lattice near $z = \frac{n\lambda}{4}$, i.e where the polarization is either $\sigma^+ or \sigma^-$ since here only one of the transition is more favoured. At these special points, the unfavoured transition becomes metastable and the electron can now be shelved (in this metastable state) hence inducing a population inversion. The ground state coherence (β) which is determined by the atomic momentum and the axial magnetic field is found to be a convenient tunable experimental parameter to induce a stimulated emission. As already indicated, the two Zeeman sublevels m= ± 1 of the $2^3S_1$ state and m=0 Zeeman sublevel of $2^3P_1$ state of helium atom form a real-life system that closely resembles our model.

## ACKNOWLEDGEMENTS:

This work was done partially within the framework of the Associateship Scheme of the Abdus Salam International Centre for Theoretical Physics, Trieste, Italy.

## REFERENCES

[1] H.J Metcalf, P Van der Straten, " Laser cooling and Trapping", (Springer, New York, 1999).

[2] A. Bhattacherjee, Opt. Comm., **191**, 83 (2001), A. Bhattacherjee, Acta Physica Slovaca, **51**, 347 (2001), A. Bhattacherjee, Journal of Optics B: Quantum and semi classical optics, **3**, 382, (2001).

[3] K.M.Gheri, H.Ritsch, D.F. Walls and V.J.Balykin, Phys. Rev. Lett., **74**, 678 (1995).

[4] P.Horak, K.M.Gheri and H.Ritsch, Phys. Rev.A **52**, 554 (1995).

[5] A. Aspect, E.Arimondo, R. Kaiser, N.Vansteenkiste and C.Cohen-Tannoudji, Phys. Rev. Lett. 61, 826 (1988); J. Opt. Soc. Am. B **6**, 2112 (1989).

[6] Ch. J. Bordi, in Advances in laser spectroscopy, edited by F.T.Arrechi, F.Strumia and H.Walter (Plenum, New York, 1983).

[7] S.Stenholm, Appl. Phys. **15**, 287 (1978).

[8]  Y. Castin, H.Wallis and J.Dalibard, J. Opt. Soc. Am.B **6**, 2046 (1989).

[9]  G.C.Hegerfeldt and M.B.Plenio, Phys. Rev. A, **52**, 3333 (1995).

[10]  B.R.Mollow, Phys. Rev. A, **5**, 1522 (1972).

[11]  E. Arimondo, in "Progress in optics", Edited by E. Wolf (Elsevier, Amsterdam, 1996), Vol. **35**, p. 257.

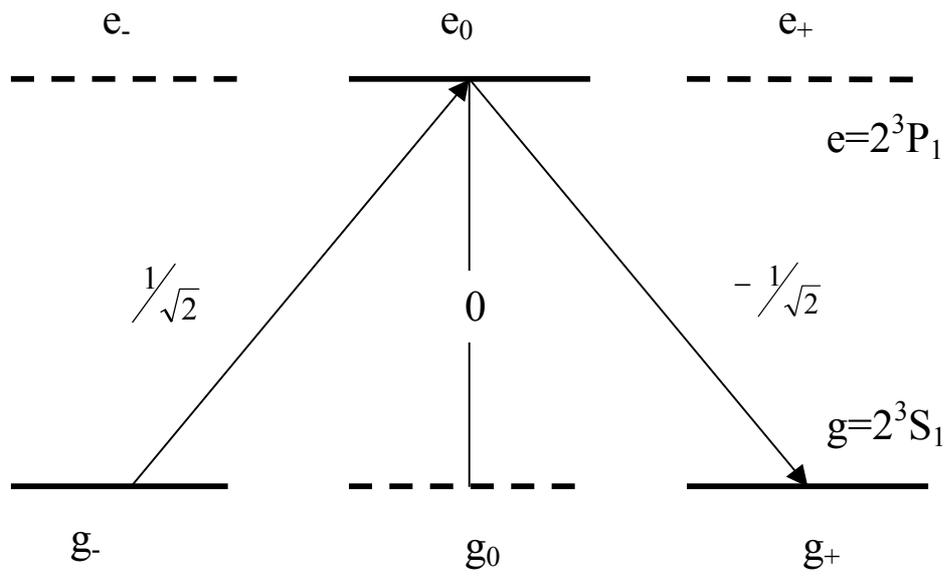

Figure 1: The Zeeman sublevels of $^4$He and the relevant Clebsch-Gordan Coefficients. Since $e_0 \leftrightarrow g_0$ transition is forbidden, all atoms are pumped into $g_+$ and $g_-$ after few fluorescence cycles. These two levels are coupled only to $e_0$, and a closed three-level $\Lambda$ configuration is realized.

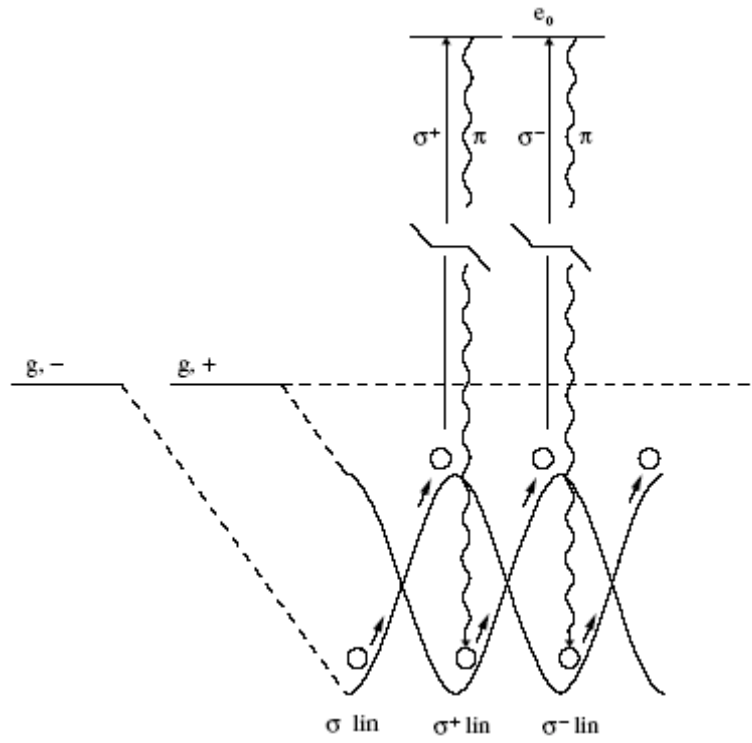

Figure 2: One-dimensional optical potential induced by interference between oppositely traveling orthogonally polarized plane waves.

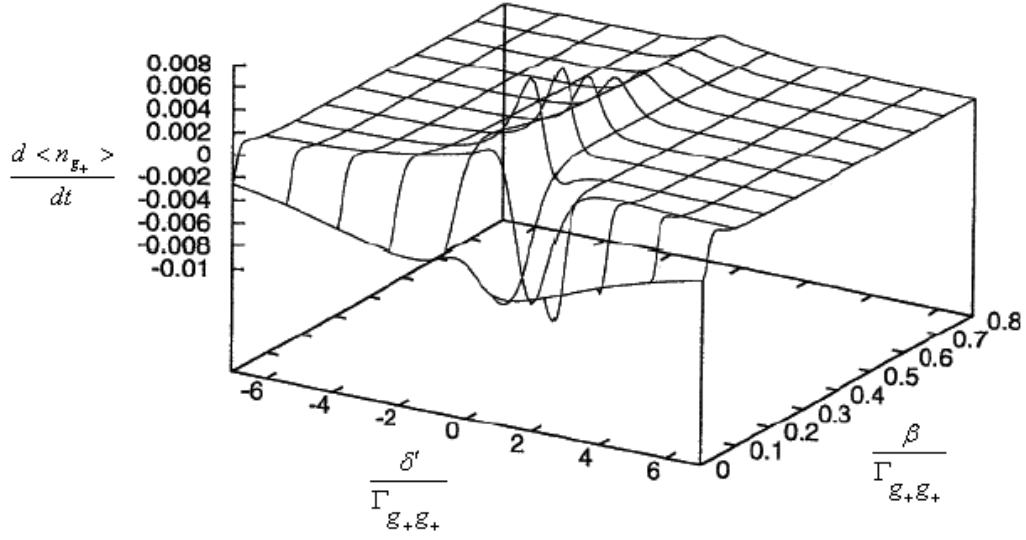

Figure 3: A three dimensional surface plot of $\dfrac{d<n_{g_+}>}{dt}$ as a function of $\dfrac{\beta}{\Gamma_{g_+g_+}}$ and $\dfrac{\delta'}{\Gamma_{g_+g_+}}$ for $\cos^2(kz)=0.1$, $\sin^2(kz)=0.9$, $\dfrac{\Omega}{\Gamma_{g_+g_+}}=10$ and $\dfrac{\Gamma_{g_-g_-}}{\Gamma_{g_+g_+}}=9$.